\documentclass[english,two column,showpacs]{revtex4}
\usepackage[T1]{fontenc}
\usepackage[latin1]{inputenc}
\usepackage{graphicx}
\usepackage{bm}
\makeatletter

\usepackage{babel}
\makeatother
\begin{document}
\def\a{\alpha}
\def\b{\beta}
\def\e{\varepsilon}
\def\d{\delta}
\def\l{\lambda}
\def\m{\mu}
\def\t{\tau}
\def\n{\nu}
\def\o{\omega}
\def\s{\sigma}
\def\S{\Sigma}
\def\G{\Gamma}
\def\D{\Delta}
\def\O{\Omega}

\def\ra{\rightarrow}
\def\ua{\uparrow}
\def\da{\downarrow}
\def\pd{\partial}
\def\bk{{\bf k}}
\def\br{{\bf r}}
\def\bm{{\bf m}}
\def\bz{{\bf z}}

\def\be{\begin{equation}}
\def\ee{\end{equation}}
\def\bea{\begin{eqnarray}}
\def\eea{\end{eqnarray}}
\def\nn{\nonumber}
\def\lb{\label}
\def\pref#1{(\ref{#1})}

\title{Quantum effects in atomically perfect specular spin valve structures}

\author{J.M.~Teixeira, J. Ventura, Yu.G. Pogorelov, and J.B. Sousa}

\affiliation{IFIMUP and Departamento de F\'{\i}sica, Universidade do
Porto, R. Campo Alegre, 687, Porto, 4169, Portugal}

\begin{abstract}
A simple tight-binding theoretical model is proposed for spin dependent,
current-in-plane transport in highly coherent spin valve structures under
specularity conditions. Using quantum-mechanically coherent and spatially
quantized Fermi states in the considered multilayered system, a system of
partial Boltzmann kinetic equations is built for relevant subbands to yield
the expressions for conductance in parallel or antiparallel spin valve
states and thus for the magneto-conductance. It is shown that specularity
favors the magnetoresistance to reach its theoretical maximum for this structure
close to $100\%$. This result is practically independent of the model parameters,
in particular it does not even need that lifetimes of majority and minority
carriers be different (as necessary for the quasiclassical regimes). The main
MR effect in the considered limit is due to the transformation of coherent
quantum states, induced by the relative rotation of magnetization in the FM
layers. Numerical calculation based on the specific Boltzmann equation with an
account of spin-dependent specular reflection at the interfaces is also performed
for a typical choice of material parameters.
\end{abstract}
\pacs{75.10.Hk; 75.30.Gw; 75.70.Cn; 76.50.+g}
\maketitle

\section{Introduction}
\begin{figure}
   \includegraphics[width=8.5 cm]{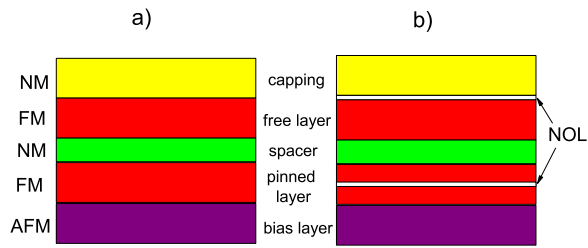}\\
  \caption{Schematics of spin valve structures: a) common
  and b) specular.}\label{spec1}
\end{figure}

Fabrication of novel nanostructured spintronics devices and the related
experimental studies of spin-dependent electron transport stimulate
new theoretical approaches to the physical properties of nanosystems where
quantum coherence effects can have a decisive role, in contrast to the mostly
quasiclassical framework of the traditional electronics. One important
class of such systems concerns spin valves \cite{dieny} formed by two ferromagnetic
(FM) layers separated by a thin non-magnetic (NM) spacer. The magnetization of
one of the FM layers (called pinned layer) is fixed by the bias from underlying
antiferromagnetic (AFM) layer, while the magnetization of the other FM layer (free layer)
easily rotates when a small magnetic field is applied. This significantly affects the
in-plane conductance, leading to relatively high MR values, typical for giant
magnetoresistance (GMR) \cite{baibich}, but the technology still demands further
improvements. One of them consists in the introduction of nano-oxide layers (NOL's)
just above the free layer and inside the pinned layer (so that the pinning is not
disrupted) \cite{NOL}. Such NOL-equipped device, the so called specular spin valve
(SSV, Fig. \ref{spec1}b) can more than double the GMR ratio of simpler stacks (Fig.
\ref{spec1}a). The increase of MR is believed to arise from the specular reflection
of electrons at the FM/NOL interfaces.

But, besides the evident effect of carriers confinement, the reduced normal-to-plane scale
$d$ of magnetic layers (few nm thickness, controlled within a 1 \AA~precision) might
allow for a pronounced quantization of the normal component of quasi-momentum, as already
indicated by the recent data on spin-resolved electronic reflection from magnetic nanolayers
\cite{zdyb}, \cite{graf}. Furthermore, it is expected that the relevant modes at the Fermi
level for each polarization are dramatically restructured when the mutual polarization of
magnetic layers is changed. All this can qualitatively change the kinetics of spin-dependent
transport, compared to the usual diffusive scenario for a quasi-continuous spectrum \cite{camley}.
However, the microscopic understanding of electron specular interface reflection is still far
from complete, in particular its role in size quantization and coherence of Fermi states. Here
we propose a theoretical description of these effects, through a properly modified Boltzmann
kinetic equation, taking into account the formation of transverse-quantized electronic subbands
and spin-dependent specular reflection at the interfaces within the simplest tight-binding model,
easy enough to advance up to numerical calculations of the MR behavior.

\section{Model}
\begin{figure}
\includegraphics[width=7.5cm]{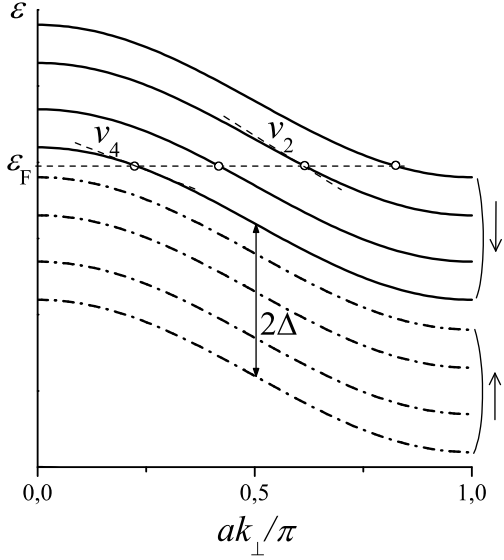}
\caption{\label{spec2}Sketch of the dispersion laws (along the diagonal
$k_x = k_y = k_\perp$ of 2D Brillouin zone) in spin-splitted and spatially
quantized subbands of a magnetic nanolayer. The circles indicate Fermi
momenta for particular (minority spin) subbands and the related Fermi
velocities $v_\a$ correspond to the slopes of dispersion laws.}
\end{figure}

Let us begin from a single metal layer, made of $n$ atomic planes with simple cubic lattice
coordination and hopping integral $t$ between nearest neighbors at distance $a$. The respective
electronic spectrum for given spin polarization $\s = \ua,\da$ and planar quasimomentum $\bk$
consists of $n$ subbands of the form $\e_{\a,\bk,\s}=\e_{\bk}+\D_{\s}+\d_{\a}$ (Fig. \ref{spec2}).
Here $\e_{\bk} = 2 t \left(2-\cos ak_{x}-\cos ak_{y}\right)$ is the 2D dispersion law for a single
plane, and in a ferromagnetic metal it is accompanied by the Stoner energy shift $\D_{\s} = \pm \D$
for minority and majority spins respectively. The spatial quantization is accounted for by the
subband shifts $\d_{\a}$ ($\a = 1,\dots,n$) which are the eigenvalues of the $n\times n$ secular
equation

\be
 \left|\begin{array}{ccccc} \delta & t & 0 & \dots & 0\\
t & \delta & t & \dots & 0\\
\dots & \dots & \dots & \dots & \dots\\
0 & \dots & t & \delta & t\\
0 & \dots & 0 & t & \delta\end{array}\right|=0
 \lb{eq:1}
 \ee

\noindent with exact values $\d_\a = 2t \cos \left[\pi\a/(n+1)\right]$. The wave function for
the $\a,\bk,\s$ state, at the planar position $\br$ in the $j$-th plane, is
$\psi_{\a, \bk,\s} (\br,j ) = A_{j}^{(\a)}{\rm e}^{i \bk \cdot \br} \chi_{\s}$, where the
components of the $n$-dimensional eigenvector $A^{(\a)}$ related to the eigenvalue $\delta_{\alpha}$
are explicitly given by

\be
 A_j^{(\a)}=\sqrt{\frac 2{n+1}}\sin\frac{\pi\a j}{n+1},
 \lb{eq:2}
  \ee

\noindent and $\chi_{\sigma}$ is the spin function.

\begin{figure}
\includegraphics[width=7.5cm]{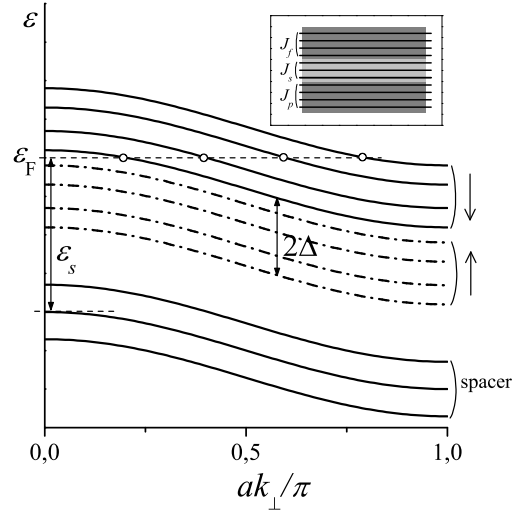}
\caption{\label{spec3}Energy band structure in the trilayered system. All
the modes are doubly degenerate and the relevant modes at the Fermi level
are marked with circles. Inset: spatial composition of atomic planes forming
the sets $J_{f,s,p}$ in \emph{f}-, \emph{s}-, and \emph{p}-layers.}
\end{figure}

Next the model is extended to include the hopping $t^{\prime}$
between the neighbor FM and NM layers, hybridizing the subbands of
the free FM layer (composed of $n_{f}$ atomic planes)
$\varepsilon_{\alpha,\mathbf{k},\sigma}^{f}$, of the NM spacer
(composed of $n_{s}$ planes) $\varepsilon_{\mathbf{\alpha,k}}^{s}$
(with $\Delta=0$), and of the pinned FM layer ($n_{p}$ planes)
$\varepsilon_{\alpha,\mathbf{k},\sigma}^{p}$. We shall denote the
respective eigenvectors (for the uncoupled layers) by $F^{(\a)},S^{(\a)}$,
and $P^{(\a)}$, with the components given again by Eq. \ref{eq:2} for
$n = n_f, n_p, n_s$, while the notation $M^{(\a)}$ is adopted for the
eigenvectors of the coupled system. The specularity effect in this approach
is modeled by zero coupling of the FM layers to their outer neighbors.
The resulting spectrum totals up to $n_t=2(n_{f}+n_s+n_{p})$ spin-resolved
modes with energies $\e_{\a,\bk}$ and wave functions $\Psi_{\a,\bk}(\br,j) =
M_j^{(\a)}{\rm e}^{i \bk \cdot \br}\chi_{\s(\a)}$, where $\a =
1,\dots,n_t$ and $\s(\a)$ is the implicit polarization of $\a$-th
mode (Fig. \ref{spec3}). We emphasize that from the total of $n_t$ modes,
only a smaller number, $n_{r}$, of modes, those present on the Fermi level,
are relevant for conductance. Thus, for the characteristic case of FM Co
layers, only minority spin subbands should take part in the transport (as
suggested by the bulk Co band structure \cite{band}). Moreover, we have to
take into account the sizeable differences in the corresponding Fermi
velocities $v_\a$ (practically coincident with those in the uncoupled layers,
Fig. \ref{spec2}). The most essential effect of hybridization is on the
amplitudes $M_j^{(\a)}$ which are generally some weighted combinations of all
the $F,P,S$ modes, and the crucial point is that the weights of $F,P$ components
in the relevant modes are strongly dependent on the mutual polarization of FM
layers (see below).

Then the kinetics of the composite system is described by the set of
$n_{r}$ distribution functions $f_{\a,\bk} = f_{\a,\bk}^{(0)} +
g_{\a,\bk}$ where $f_{\a,\bk}^{(0)} = \left[{ \rm e}^{\b
\left(\e_{\alpha,\bk}-\e_{\rm F}\right)} + 1 \right]^{ -1}$ is the
usual equilibrium Fermi function with $\b = 1/k_{\rm{B}}T$ and
$g_{\a,\bk}$ is the non-equilibrium part due to the external electric
field $\mathbf{E}$. The current density is given by the sum

\be
 \mathbf{j}=\frac e {n a}{\sum_{\alpha}}^\prime\int\frac{d\mathbf{k}}
  {\left(2\pi\right)^{2}} \mathbf{v}_{\a,\bk} g_{\a,\bk},
   \label{eq:3}
    \ee

\noindent where $\sum^\prime$ means summation over the $n_r$ relevant
modes, $\mathbf{v}_{\a,\bk} = \hbar^{-1}\pd \e_{\a,\bk}/\pd\bk$ is
the electron velocity, and the components of the non-equilibrium
distribution are defined from the system of Boltzmann equations:

\be
 \frac{e\mathbf{E}}{\hbar}\cdot\frac{\partial f_{\a,\bk}^{(0)}}{\partial\bk}
  + {\sum_{\b}}^{\prime\prime}  \int \frac{a^2 d\bk^\prime}{(2\pi)^2}
   \o_{\a,\bk}^{\b,\bk^\prime} \left(g_{\b,\bk^\prime} -
    g_{\a,\bk}\right)= 0.
     \lb{eq:4}
      \ee

\begin{figure}
\includegraphics[width=8cm]{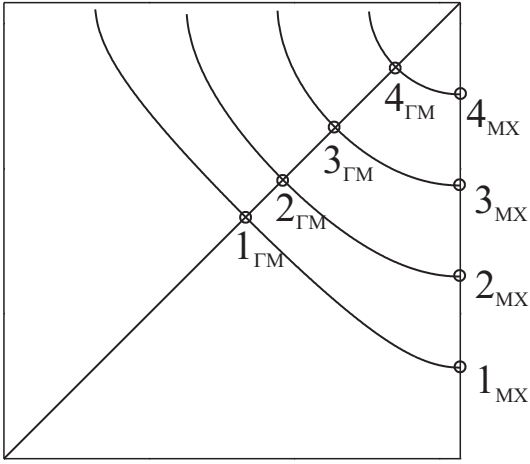}
\caption{\label{spec4}Configurations of Fermi lines for spatially
quantized subbands of minority electrons in the Brillouin zone. The
characteristic points along high symmetry directions $\G M$ and $MX$
were used to approximate the averages of $v_\a^{-1}$ and $v_\a^2$.}
\end{figure}

\noindent Here $\sum^{\prime\prime}$ means summation over relevant
modes with conserved spin, $\s(\a)=\s(\b)$, and $\o_{\a,\bk}^{\b,\bk^\prime}$
is the transition rate due to scattering from $\bk$ state of $\a$-th subband
to $\bk^\prime$ state of $\b$-th subband. We consider transitions only due
to elastic scattering by random point-like impurities with potential $V$ and
concentration $c \ll 1$ (per unit cell). Then the Fermi Golden Rule
transition rates are $\omega_{\a,\bk}^{\b,\bk^\prime}= \O_{\a,\b}
\delta\left(\e_{\a,\bk} -\e_{\b, \bk^\prime} \right)$ with the
scattering factors (averaged in impurity positions)

\be
 \O_{\a,\b}=\frac{2\pi cV^{2}}{\hbar n}\sum_{j}\left|M_j^{(\a)}
  M_{j}^{\left(\b\right)}\right|^{2}.
   \lb{eq:5}
    \ee

\noindent In this simple model, the first term in the collision
integral of Eq. \ref{eq:4} turns out to be proportional to $\int d\bk
 g_{\b, \bk}\d\left(\e_{\rm F} - \e_{ \b,\bk} \right)$, that is to
the average of the non-equilibrium distribution over the Fermi surface
and so it should vanish. Then the solution takes the common form
$g_{\a,\bk} = \hbar^{-1}\t_\a e\mathbf{E} \cdot \partial
f_{\a,\bk}^{(0)}/\partial\bk$ where the relaxation time for the
$\a$-th mode is defined by

\be
 \t_\a^{-1} = {\sum_{\b}}^{\prime\prime} \rho_{\b}\O_{\a,\b},
  \lb{eq:6}
   \ee

\noindent including the Fermi densitiy of states $\rho_{\b} = (a/2
\pi)^2 \int d\bk \d\left(\e_{\b,\bk} - \e_{ \rm{F}} \right)$ for
each $\b$-th mode. Then the total conductivity is found from Eq.
\ref{eq:3} as a sum of partial contributions:

\be
 \s_{tot} = {\sum_\a}^\prime \s_\a, \quad \s_\a = \frac{e^2 \t_\a
  \rho_{\a} \left\langle v_{\a}^2\right\rangle}{n a^3},
   \ee

\noindent where $\left\langle v_{\a}^2\right\rangle \approx
\rho_{\a}^{-1}\left(a/2\pi\right)^2\int d\bk v_{\a,\bk}^2
\d\left(\e_{\a,\bk} - \e_{\rm{F}} \right)$ is the average of the
respective squared Fermi velocity. In fact, this is a particular
case of the general Landauer formula \cite{land}, written for the
present system of $n_r$ coherent quantum channels.

The system, Eqs. \ref{eq:1}-\ref{eq:6}, can be routinely treated by
numerical methods at any relative orientation of magnetizations in
$f$- and $p$- layers, from parallel ($\ua\ua$) to antiparallel
($\ua\da$), to result in the principal quantity of interest, the
magnetoresistance

\be
 \frac{\D R}R = \frac{\s_{tot}^{\ua\ua}}{\s_{tot}^{{\ua\da}}}-1 =
 \frac{{\sum_\a}^\prime\rho_\a\left\langle v_\a^2\right\rangle\t_\a^{\ua\ua}}
  {{\sum_\a}^\prime\rho_\a\left\langle v_\a^2\right\rangle\t_\a^{\ua\da}}-1.
   \lb{eq:7}
    \ee

But some qualitative conclusions about the specularity effect on MR
in a nanolayered device can be drawn already from simple inspection
of the discrete structure of the amplitudes $M_j^{(\a)}$, according to
the following remarks.

First of all, we suppose that in absence of hybridization the
majority and minority subbands are well separated from each other
and from the spacer subbands (like the situation in bulk Co and Cu).
Then we notice that the $j$-configurations of the above amplitudes
are essentially different for $\ua\ua$ and $\ua\da$ cases and hence
consider them separately. Finally, an important factor for the very
existence of GMR (in this quantum coherent conductance regime) is
the presence of certain "resonances" between relevant modes at the
Fermi level. Namely, a resonance appears between two (unhybridized)
modes $\e_{\a,\bk,\s}^f$ and $\e_{\b,\bk,\s}^p$ if their energy
separation near the Fermi level is less then the effective coupling
$\sim {t^\prime}^2/\e_s $ (mediated by the spacer modes at typical
energy distance $\e_s$, see Fig. \ref{spec3}). Moreover, for the
sake of clarity, we shall restrict the following consideration to
the simplest situation of identical \emph{f}- and \emph{p}-layers
where all $n_f = n_p$ modes are relevant and can form resonant
$fp$-pairs.

Thus, in the $\ua\ua$ configuration, there appears a strong
hybridization in each $F^\a$, $P^\a$ pair, forming two collective
modes as their bonding and anti-bonding combinations (in neglect of
small contributions $\sim {t^\prime}^2/\left(\e_s\D\right) \ll 1$ of
the rest of the modes):

\be
 M_j^{\a,\pm} \approx \frac 1 {\sqrt 2} \left\{
  \begin{array}{c} F_j^\a,\quad{\rm for}\quad j \in J_f, \\
   0,\quad{\rm for}\quad  j\in J_s, \\
   \pm P_j^\a,\quad{\rm for}\quad  j \in J_p,
    \end{array}\right.
     \lb{eq:8}
     \ee

\noindent where $J_{f,s,p}$ are the sets of atomic planes entering
\emph{f}-, \emph{s}-, and \emph{p}-layers (see inset in Fig.
\ref{spec3}). The respective relaxation times are given by

\bea
 &&\left(\t_{\a,\pm}^{\ua\ua}\right)^{-1} \approx \frac{\pi c V^2}{2\hbar n}
  {\sum_\b}^{\prime\prime}\rho_\b\nonumber\\
    &&\quad \times  \left(\sum_{j \in n_f}\left|F_j^\a F_j^\b\right|^2
  +  \sum_{j \in n_p}\left|P_j^\a
     P_j^\b\right|^2\right).
       \lb{eq:9}
        \eea

\noindent Then we can use the exact sum rule for the amplitudes, Eq.
\ref{eq:2}:

\be
 \sum_{j=1}^n \left( A_j^\a A_j^{\b} \right)^2 =\frac 1
  {n+1} \left(1 + \frac{\d_{\a,\b} + \d_{\a,n+1-\b} }
   2\right),
    \lb{eq:10}
     \ee

\noindent to present the relaxation times, Eq. \ref{eq:9}, as

\be
 \t_{\a,\pm}^{\ua\ua} \approx \frac{\hbar n}{2 \pi c V^2
 {\sum_\b}^{\prime\prime}\rho_\b}.
  \lb{eq:11}
  \ee

\noindent Contrariwise, in the $\ua\da$ configuration, all the
relevant modes remain almost unhybridized, taking nearly "local"
forms:

\bea
 M_j^{\a,f} & \approx &  \left\{
  \begin{array}{c} F_j^\a,\quad\quad{\rm for} \quad j\in J_f, \\
   0,\quad{\rm for}\quad  j \in J_s \cup J_p,
    \end{array}\right.\nonumber\\
     M_j^{\a,p} & \approx &  \left\{
  \begin{array}{c} 0,\quad{\rm for} \quad j\in J_f\cup J_s, \\
   P_j^\a,\quad\quad{\rm for}\quad  j \in   J_p,
    \end{array}\right.
     \lb{eq:12}
     \eea

\noindent and in this approximation we obtain for the relaxation
times $\t_{\a,i}^{\ua\da}$ half the value of Eq. \ref{eq:11}. Then
the magnetoresistance, Eq. \ref{eq:7}, is readily estimated as $\D
R/R \approx 100\%$. We notice that this result is practically
independent of the parameters of interlayer coupling and impurity
scattering, in particular it does not even need that lifetimes of
majority and minority carriers be different (as necessary for
quasiclassical regimes). The main MR effect in the considered limit
is due to the variation of coherent quantum states, induced by the
relative rotation of magnetization of the FM layers.

\section{Numerical calculations}

\begin{table}
    \begin{center}
        \begin{tabular}{lcccc}
        \hline \hline
    & & $10^{-7}\times\O_{\a,\b}$  (W) \\\hline
    & $\b = 1$ & $\b = 2$ & $\b = 3$ & $\b = 4$ \\
$\a = 1$ $\uparrow\uparrow$ & $0.5175$ & $0.3431$ & $0.3417$ & $0.5151$ \\
\qquad $\;$\, $\uparrow\downarrow$ & $1.0347$ & $0.6857$ & $0.6824$ & $1.0287$ \\
                                                                       \\
$\a = 2$ $\uparrow\uparrow$ & $0.3431$ & $0.5118$ & $0.5097$ & $0.3415$ \\
\qquad $\;$\, $\uparrow\downarrow$ & $0.6857$ & $1.0224$ & $1.0175$ & $0.6818$ \\
                                                                        \\
$\a = 3$ $\uparrow\uparrow$ & $0.3417$ & $0.5097$ & $0.5077$ & $0.3402$ \\
\qquad $\;$\, $\uparrow\downarrow$ & $0.6824$ & $1.0175$ & $1.0127$ & $0.6786$ \\
                                                                        \\
$\a = 4$ $\uparrow\uparrow$ & $0.5151$ & $0.3415$ & $0.3402$ & $0.5127$ \\
\qquad $\;$\, $\uparrow\downarrow$ & $1.0287$ & $0.6818$ & $0.6786$ & $1.0228$ \\
\hline \hline
        \end{tabular}
    \end{center}
    \caption{Scattering factors ($\O_{\a,\b}$) for the Fermi modes ($\a, \b = 1$,
    $2$, $3$ and $4$) in the parallel ($\uparrow\uparrow$) and antiparallel
    ($\uparrow\downarrow$) configurations.}
    \label{table1}
\end{table}

\begin{table}
    \begin{center}
        \begin{tabular}{lcccc}
        \hline \hline
$\a$ & 1 & 2 & 3 & 4 \\\hline
$10^{19}\rho_{\a}(\texttt{J}^{-1})$ & $0.209$ & $0.2439$ & $0.3238$ & $0.5517$ \\
$10^{10}\langle v_{\a}^{2}\rangle (\texttt{m}^{2}/\texttt{s}^{2})$ & $0.7431$ & $2.2595$ & $2.9334$ & $2.6008$ \\
$10^{-12}\tau_{\a}^{\uparrow\uparrow}(\texttt{s})$ & $1.7018$ & $1.8157$ & $1.823$ & $1.7094$ \\
$10^{-12}\tau_{\a}^{\uparrow\downarrow}(\texttt{s})$ & $0.8515$ & $0.9091$ & $0.9133$ & $0.8563$ \\
\hline \hline
        \end{tabular}
    \end{center}
    \caption{The Fermi density of states $\rho_{\a}$, averages of squared Fermi
    velocities $\langle v_{\a}^{2}\rangle$ and relaxation times $\t_\a$ for the Fermi modes
    ($\a = 1$, $2$, $3$ and $4$) in the parallel ($\uparrow\uparrow$) and antiparallel
    ($\uparrow\downarrow$) configurations.}\label{table2}
\end{table}

To certify the above qualitative considerations, a detailed
numerical calculation was done for the particular choice of
parameters: $t = t^\prime = 0.25$ eV, $\D = 0.5$ eV, $\e_s = 2$ eV
(a single-band model for real \emph{d}-bands of Co and Cu), $n_f =
n_p = 4$, $n_s = 3$ (a simple discrete structure of layers), and $V
= 0.5$ eV, and $c = 0.01$ (typical impurity parameters). The Fermi
velocities (and their inverse values) for two characteristic
directions in Brillouin zone were used to approximate the partial
densities of states

\[\rho_\a \approx \frac{a^2L_\a}{8\pi2\hbar}\left( \frac 1{v_{\a,\G M}}
+ \frac 1 {v_{ \a, MX}}\right),\]

\noindent ($L_\a$ being the length of respective Fermi line, Fig.
\ref{spec4}), and then $\langle v_\a^2\rangle \approx v_{\a,\G M}v_{
\a, MX}$. The obtained numerical results for $\O_{\a,\b}$,
$\rho_\a$, $\t_\a$, and $\langle v_\a^2 \rangle$ are illustrated in
tables \ref{table1} and \ref{table2}, respectively, for the $\ua\ua$ and
$\ua\da$ configurations. These numerical values lead to $\approx
99.65$\% of magnetoresistance, that is quite close to the maximum
possible MR = 100\% in the coherent regime. To compare with, for purely
incoherent currents there will be no MR at all in such two-layer
system, so that the finite effect only appears from their partial
mixing due to scattering at the interfaces \cite{camley} and is
estimated phenomenologically as $\sim\exp\left(-d/\ell\right)$ of
the above maximum value.

Actually, the experimental MR values in specular spin valves
\cite{gehanno,sousa} are clearly lower than the above model estimates.
This can be due to a number of important factors, not included into
the present simple model (which therefore should be considered as a
certain idealized reference case). First of all, the postulated ideal
specularity condition (supposing the wave function fully confined
within the $n$-plane system) cannot be exact in reality, and a
considerable part of the electronic density can "escape" through the
NOL barriers to adjacent non-magnetic (or AFM) layers. This part would
act as a parallel conduction channel, mostly unchanged at reorientation
of FM electrodes and hence restricting the magnetoresistive effect.
Also, the used model of rigid Stoner shifts of spin subbands in FM
electrodes of course overestimates sharpness of the spin-dependent
energy barrier between these electrodes, where in fact the band
structures are not uniform on scales of few atomic layers. Other
restrictive factors are the temperature effect (by phonons and magnons),
the roughness in the FM/NM interfaces and the presence of defects as
grain boundaries, displacements and distortions in the crystalline
structure, which all reduce the coherence of relevant quantum states
and so the validity of the Landauer formula. At last, the single-band
model may be oversimplified, compared to the real hybridized \emph{s-d}
band structures of transition metals used in various numerical
studies of spin-valves \cite{tsymbal1,Chen,Blaas}, but it would
be rather problematic to keep the analytic description at such
detailed level. Nevertheless, a further development within the
present model can be done, varying the number $n_r$ of relevant
modes and admitting the presence of both spin polarizations among
these modes.

\section{Conclusions}

A simple single-band tight-binding model was developed to estimate
theoretically the maximum possible enhancement of GMR in the system of
quantum coherent FM nanolayers, using a specific set of Boltzmann equations
for spatially quantized and spin-resolved subbands and a Landauer-type
formula for the spin-dependent conductance. It is shown that a limit GMR
value close to 100\% can be reached for a fully coherent (and fully specular)
SSV nanostructure and the reducing factors for this value in real SSV systems
are discussed.

\begin{acknowledgments}
This work was supported in part by FEDER-POCTI/0155, POCTI/CTM/45252/02 and POCTI/CTM/59318/2004 from Portuguese FCT and IST-2001-37334 NEXT MRAM projects.
JMT and JV are thankful for FCT grants (SFRN/BD/24012/2005 and SFRN/BPD/2163/2005).
\end{acknowledgments}

\end{document}